\documentclass[conference]{IEEEtran}

\usepackage{mathtools}
\usepackage{amssymb}
\usepackage{amsmath}
\usepackage{algorithm}
\usepackage{enumerate,multirow}
\usepackage{algpseudocode}
\usepackage{tikz}
\usepackage{tikz}
\usetikzlibrary{patterns,positioning,arrows,calc}
\usepackage{siunitx}
\usepackage{multirow}
\usepackage{multicol}
\usepackage{color}
\usepackage{xcolor}
\usepackage{algorithm}
\usepackage{algpseudocode}
\usepackage{graphicx}
\usepackage{subcaption}
\usepackage{booktabs}

\usepackage{comment}

\usepackage{geometry}
\usepackage{pgfplots}   
\pgfplotsset{compat=1.18}
\usepackage{tikz}
\usepackage{tikz}
\usetikzlibrary{patterns,positioning,arrows,calc}\usepackage{graphicx}
\usepackage{graphicx}
\usepackage{tikz}
\usetikzlibrary{trees}
\date{}

\usepackage{calc}
\newcolumntype{M}[1]{>{\centering\arraybackslash}m{#1}}
\newcolumntype{N}{@{}m{0pt}@{}}

\usepackage{mdwmath}
\usepackage{blindtext}
\usepackage{eqparbox}
\usepackage{stfloats}
\usepackage[numbers,sort&compress]{natbib}
\usepackage{algorithm}
\usepackage{algpseudocode}
\usepackage{tikz}
\usetikzlibrary{patterns, arrows.meta, positioning}

\IEEEoverridecommandlockouts
\title{\Huge
Logically Augmented Quantum LDPC Codes: Construction and Hierarchical Decoding}
\author{
\IEEEauthorblockN{
Sheida Rabeti$^{1}$,
Ayushi Shirke$^{2}$,
Ivana Dimitrova$^{1,2}$,
and Hessam Mahdavifar$^{1}$
}
\IEEEauthorblockA{
$^{1}$Department of Electrical and Computer Engineering, Northeastern University, Boston, MA 02115, USA\\
$^{2}$Department of Physics, Northeastern University, Boston, MA 02115, USA\\
Email: \{rabeti.s, shirke.a, i.dimitrova, h.mahdavifar\}@northeastern.edu
}
\thanks{This work was supported by the U.S. Department of Energy under grant DE-SC0026502.}
}

\newtheorem{example}{{Example}}

\newtheorem{definition}{{Definition}}
\newtheorem{remark}{{Remark}}

\DeclareMathAlphabet{\mathbfsl}{OT1}{ppl}{b}{it} 

\newcommand{\be}[1]{\begin{equation}\label{#1}}
\newcommand{\ee}{\end{equation}}

\renewcommand{\leq}{\leqslant}

\newcommand{\Cref}[1]{Co\-ro\-lla\-ry\,\ref{#1}}

\usepackage{hvlogos}
\usepackage{qtree}
\usepackage{tikz}
\usepackage{multirow}
\usepackage{hhline}
\usepackage{hyperref}
\hypersetup{hidelinks}

\begin{document}

\vspace{10mm}
\maketitle

\begin{abstract}
We propose a joint framework for constructing and decoding Logically Augmented (LA) quantum low-density parity-check (QLDPC) codes. This is motivated by the need for flexible coding rates and adaptive error protection while preserving block length and code structure. On the construction side, we use an auxiliary \textit{Logical Protection Code (LPC)} to impose additional stabilizer constraints on selected logical qubits of a \textit{primary} code. On the decoding side, the proposed decoder uses the LPC syndrome to generate logical error hypotheses and condition belief propagation (BP) on each hypothesis. The proposed \textit{Hierarchical Logical List Decoder (HLLD)} is invoked only when the initial BP decoder for primary code fails to converge. It then selects the correction with the smallest Hamming weight among the syndrome-consistent candidates produced by hypothesis-conditioned BP runs.

For numerical evaluation, we focus on short repetition LPCs, which provide a flexible means of trading rate for additional error protection. Each LPC syndrome admits exactly two logical error hypotheses, irrespective of repetition length, which enables HLLD to explore both possibilities without relying on a single logical estimate. Numerical results show that varying the repetition length provides a tunable trade-off between rate and error-correction performance across a wide range of primary code constructions and their operating regimes. On the decoding side, with the overall LA code held fixed, HLLD reduces the logical error rate by up to 81\%, 75\%, and 69\% for LA codes based on mitten, lifted-product, and bivariate bicycle codes, respectively, compared with conventional BP decoding of the same LA codes.
\end{abstract}

\section{Introduction}
\label{sec:introduction}

Quantum error-correcting codes are essential for fault-tolerant quantum computation. Among existing constructions, quantum low-density parity-check (QLDPC) codes, particularly those of Calderbank-Shor-Steane (CSS) form, have attracted significant attention due to their sparsity constraints. In particular, these constraints reduce qubit-qubit interactions during syndrome extraction and make QLDPC codes well suited for architectures with limited qubit connectivity \cite{7336474, Breuckmann_2021,vasic2025quantumlowdensityparitycheckcodes}. Many practical finite-length QLDPC codes are obtained from algebraic or product-based constructions through extensive numerical searches. Although these frameworks support a range of code rates, adjusting the rate and error-protection profile while maintaining a given number of physical qubits and primary code structure can be challenging. Moreover, many commonly used codes are designed to provide balanced protection against Pauli-$X$ and Pauli-$Z$ errors, whereas physical noise may exhibit substantially different error rates for these error types \cite{evered2023high, evered2026high}. These limitations call for a flexible framework that adapts the rate and error protection of existing QLDPC codes to different noise regimes and target logical error rates, while preserving the physical block length and overall code structure.

To address these limitations, we propose a framework for constructing Logically Augmented (LA) QLDPC codes. In this construction, an auxiliary CSS code, termed the \textit{Logical Protection Code (LPC)}, imposes additional stabilizer constraints on selected logical qubits within a single primary QLDPC code block. The choice of LPC enables a trade-off between rate and error protection, while allowing the relative protection against Pauli-$X$ and Pauli-$Z$ errors to be tailored to the noise characteristics. Here, the physical block length and all original stabilizer generators of the underlying \textit{primary} code are preserved. This approach is related to quantum concatenation and generalized concatenation \cite{knill1996concatenated,grassl2009generalized}, as well as code expurgation, which converts selected logical operators into additional stabilizers \cite{gullans2021quantum}. More recently, yoked surface codes have introduced parity constraints across encoded surface-code blocks \cite{gidney2025yoked}. Also, elevator codes combine repetition inner codes with high-rate outer codes to reduce physical-qubit overhead under biased noise \cite{shanahan2026elevator}. The main novel aspect of our construction is a structured LPC-based augmentation that operates entirely within a single QLDPC block, without increasing the physical block length or coupling separately encoded blocks, while preserving all original stabilizer generators.

On the decoding side, belief propagation (BP) offers a low-complexity approach for decoding QLDPC codes, with numerous extensions based on post-processing, decimation, stabilizer inactivation, and ensemble or list methods \cite{panteleev2021degenerate,yao2024belief,du2022stabilizer,koutsioumpas2025automorphism,rabeti2026multiple}. Related work on concatenated codes exploits higher-level syndrome information through hierarchical decoding \cite{poulin2006optimal,zhang2026bidirectional,kakizaki2026approximate}. Our proposed decoder and its numerical evaluation utilize BP and BP with ordered-statistics decoding (BP-OSD) as constituent decoders for the underlying primary code. More specifically, we propose the Hierarchical Logical List Decoder (HLLD) as an integral part of the LA framework. When initial BP decoding of the primary code fails to converge, HLLD uses the LPC syndrome to generate admissible logical error hypotheses and conditions a separate BP run on each hypothesis. It then selects the correction with the smallest Hamming weight among the resulting syndrome-consistent candidates. This approach leverages a fundamental coupling between construction and decoding: the LPC determines both the additional stabilizer constraints and the logical hypotheses that organize the decoding search. For numerical evaluation, we focus on LA codes with repetition LPCs, whose adjustable length provides a simple way to explore the rate–protection trade-off. Another advantage of this choice is that the exhaustive logical list contains exactly two hypotheses, irrespective of repetition length. We then compare HLLD with direct decoding of the same logically augmented code, explicitly demonstrating the benefit of exploiting the
construction’s hierarchy.

The rest of this paper is organized as follows.
In Section II, some preliminaries on CSS codes
and BP decoding are provided. In Section III,
we present the proposed LA CSS code construction.
Section IV describes the HLLD framework,
detailing the decoding procedure, logical-list
generation strategies, and complexity analysis.
Numerical results and performance comparisons
are presented in Section V. Finally, we conclude the paper in Section VI.

\section{Preliminaries}
\subsection{Quantum Stabilizer and CSS Codes}
An $[[N, k, d]]$ quantum stabilizer code is a $2^k$-dimensional subspace $\mathcal{C} \subseteq (\mathbb{C}_2)^{\otimes N}$ with an Abelian stabilizer group 
$\mathcal{S}$:
\begin{equation} \label{eq:StabilizerCode} \mathcal{C} = \{|\psi\rangle \in (\mathbb{C}_2)^{\otimes N} \colon s |\psi\rangle= |\psi\rangle, \forall s \in \mathcal{S} \}. \end{equation}
Each generator $g \in \mathcal{S}$ acts as a parity-check constraint. The \textit{minimum distance} d is the minimum weight of a Pauli operator that commutes with every
stabilizer and acts nontrivially on the encoded qubits. A Calderbank--Shor--Steane (CSS) code is a stabilizer code with a parity-check matrix of the form
$
H = \begin{bmatrix} H_X & 0 \\ 0 & H_Z \end{bmatrix},
$
where $H_X,H_Z$ are classical binary parity-check matrices satisfying $H_X H_Z^T = 0$. Such codes can correct Pauli-$X$ and Pauli-$Z$ errors independently using $H_Z$ and $H_X$, respectively. In this work, we focus on QLDPC CSS codes, where both $H_X$ and $H_Z$ are sparse.
For CSS codes, the minimum distance admits an equivalent characterization in terms of the underlying classical codes. Specifically, the $X$-distance is defined as
\[
d_X = \min \{ \mathrm{wt}(v) : v \in \ker(H_Z)\setminus \operatorname{rs}(H_X) \},
\]
and the $Z$-distance is also defined in the same fashion. 
The minimum distance of the CSS code is given by $d=\min(d_X,d_Z)$. Vectors in $\ker(H_Z)$ correspond to Pauli-$X$ operators that commute with all stabilizers, while vectors in $\operatorname{rs}(H_X)$ correspond to X-type
stabilizers. Hence, elements of $\ker(H_Z)\setminus \operatorname{rs}(H_X)$ represent nontrivial logical $X$ operators. Similarly, $\ker(H_X)\setminus\operatorname{rs}(H_Z)$ characterizes nontrivial logical $Z$ operators,
where $\ker(\cdot)$ and $\operatorname{rs}(\cdot)$
denote the null and row spaces, respectively.
\subsection{Belief Propagation (BP) Decoding}
BP decoding operates on the Tanner graph
$\mathcal G=(V_c\cup V_v,E)$, where $V_c$ and
$V_v$ denote the check and variable nodes,
respectively, and $E$ is the edge set.
Let $d_c$ and $d_v$ denote the maximum check-node
and variable-node degrees, respectively.
For each variable $v_i$, the log-likelihood
ratio (LLR) is initialized as
$\mu_i=\log\frac{1-p}{p}$, where $p$ denotes
the physical error probability.
During decoding, check-to-variable and
variable-to-check messages are iteratively
updated using the sum-product algorithm.
For numerical stability, variable-to-check
messages are clipped to $[-K,K]$, with $K=25$
\cite{kudekar2014effect}.
Hard decisions are made according to the signs
of the posterior LLRs. Decoding terminates
when the estimated error satisfies the measured
syndrome or the maximum number of iterations
is reached.

\section{Logically Augmented (LA) CSS Codes}
\label{sec:RPC-LD}

In this section, we introduce the construction of Logically Augmented (LA) CSS codes. The main idea is to impose additional stabilizer constraints on $n_{\mathrm L}$ selected logical qubits of a primary off-the-shelf CSS code using an auxiliary Logical Protection Code (LPC), in such a way that these qubits collectively encode $k_{\mathrm L}$ logical qubits. The remaining $k-n_{\mathrm L}$ logical qubits of the primary code are left unconstrained by the LPC. These additional stabilizers introduce logical-level redundancy without increasing the physical block length or modifying the original stabilizer generators. This enables a flexible trade-off between encoding rate and error protection. Rather than relying on a separate outer error-correction layer as in conventional concatenation, our emphasis is on introducing a \textit{small} rate reduction to obtain additional syndrome information that can bootstrap the primary-code decoder when it fails to converge.

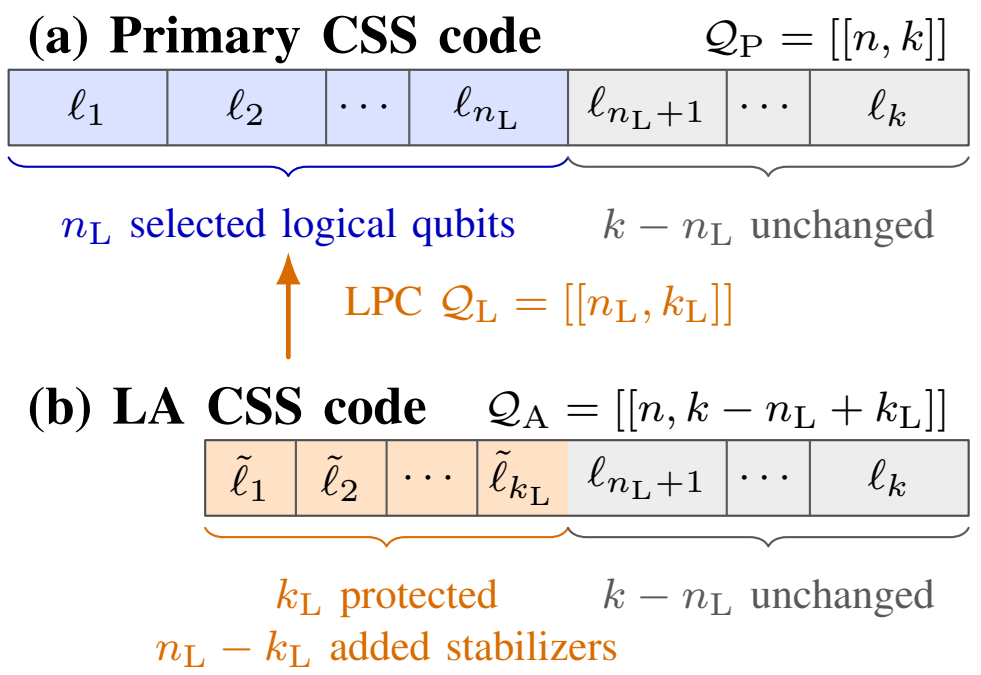
\begin{figure}[t]
\centering
\begin{tikzpicture}[
    x=1cm,y=1cm,
    font=\footnotesize,
    >=Latex,
    line cap=round,
    line join=round
]

\definecolor{selectedblue}{RGB}{218,226,255}
\definecolor{protectedorange}{RGB}{255,226,198}


\node[anchor=west,font=\bfseries]
    at (0,4.15) {(a) Primary CSS code};

\node[anchor=east]
    at (6.35,4.15)
    {$\mathcal Q_{\mathrm P}=[[n,k]]$};

\fill[selectedblue]
    (0,3.45) rectangle (3.70,3.95);

\fill[gray!14]
    (3.70,3.45) rectangle (6.35,3.95);

\draw[black!70,thin]
    (0,3.45) rectangle (6.35,3.95);

\foreach \x in {1.05,2.10,2.65,3.70,4.75,5.30}
    \draw[black!65,thin]
        (\x,3.45) -- (\x,3.95);

\node at (0.525,3.70) {$\ell_1$};
\node at (1.575,3.70) {$\ell_2$};
\node at (2.375,3.70) {$\cdots$};
\node at (3.175,3.70) {$\ell_{n_{\mathrm L}}$};

\node at (4.225,3.70) {$\ell_{n_{\mathrm L}+1}$};
\node at (5.025,3.70) {$\cdots$};
\node at (5.825,3.70) {$\ell_k$};

\draw[decorate,
 decoration={brace,mirror,amplitude=3.5pt},
 blue!75!black,thin]
 (0,3.37) -- (3.70,3.37)
 node[midway,below=6pt]
 {$n_{\mathrm L}$ selected logical qubits};

\draw[decorate,
 decoration={brace,mirror,amplitude=3.5pt},
 black!65,thin]
 (3.70,3.37) -- (6.35,3.37)
 node[midway,below=6pt]
 {$k-n_{\mathrm L}$ unchanged};


\draw[-{Latex[length=2.2mm]},
      orange!85!black,
      line width=0.9pt]
    (1.85,2.05) -- (1.85,2.75);

\node[
    anchor=west,
    text=orange!85!black
] at (2.10,2.40)
    {LPC $\mathcal Q_{\mathrm L}=[[n_{\mathrm L},k_{\mathrm L}]]$};


\node[anchor=west,font=\bfseries]
    at (0,1.70) {(b) LA CSS code};

\node[anchor=east]
    at (6.35,1.70)
    {$\mathcal Q_{\mathrm A}=[[n,k-n_{\mathrm L}+k_{\mathrm L}]]$};



\def\xOrange{1.30}
\def\xGray{3.70}
\def\xEnd{6.35}

\fill[protectedorange]
    (\xOrange,1.00) rectangle (\xGray,1.50);

\fill[gray!14]
    (\xGray,1.00) rectangle (\xEnd,1.50);

\draw[black!70,thin]
    (\xOrange,1.00) rectangle (\xEnd,1.50);

\foreach \x in {1.90,2.50,3.10}
    \draw[black!65,thin]
        (\x,1.00) -- (\x,1.50);

\foreach \x in {4.75,5.30}
    \draw[black!65,thin]
        (\x,1.00) -- (\x,1.50);

\node at (1.60,1.25) {$\tilde{\ell}_1$};
\node at (2.20,1.25) {$\tilde{\ell}_2$};
\node at (2.80,1.25) {$\cdots$};
\node at (3.40,1.25) {$\tilde{\ell}_{k_{\mathrm L}}$};

\node at (4.225,1.25) {$\ell_{n_{\mathrm L}+1}$};
\node at (5.025,1.25) {$\cdots$};
\node at (5.825,1.25) {$\ell_k$};

\draw[
    decorate,
    decoration={brace,mirror,amplitude=3.5pt},
    orange!85!black,thin
]
(\xOrange,0.92) -- (\xGray,0.92)
node[midway,below=6pt]
{$k_{\mathrm L}$ protected};

\draw[
    decorate,
    decoration={brace,mirror,amplitude=3.5pt},
    black!65,thin
]
(\xGray,0.92) -- (\xEnd,0.92)
node[midway,below=6pt]
{$k-n_{\mathrm L}$ unchanged};

\node[
    text=orange!85!black,
    font=\footnotesize
] at (2.50,0.12)
{$n_{\mathrm L}-k_{\mathrm L}$ added stabilizers};


\end{tikzpicture}
\vspace{-2mm}
\caption{Logical-level encoding of the LA CSS code
using the LPC $\mathcal Q_{\mathrm L}$ and
the primary code $\mathcal Q_{\mathrm P}$.}
\label{fig:LA_general}
\vspace{-2mm}

\end{figure}

Fig. \ref{fig:LA_general} demonstrates a block diagram for the proposed
LA-CSS construction, which we formally define next. 


\begin{definition}[Logically Augmented (LA) CSS Code]
Let $\mathcal{Q}_{\mathrm P}=[[n,k,d]]$ be a
\emph{primary CSS code} defined by the parity-check
matrices
$H_X\in\mathbb{F}_2^{r_X\times n},\ 
H_Z\in\mathbb{F}_2^{r_Z\times n}, $
satisfying
$H_XH_Z^T=0.
$

Let $L_X,L_Z\in\mathbb{F}_2^{n_{\mathrm L}\times n}$ denote
physical representatives of $n_{\mathrm L} \leq k$ selected
pairs of independent logical operators of
$\mathcal{Q}_{\mathrm P}$, satisfying
$H_ZL_X^T=0,\ 
H_XL_Z^T=0,\ 
L_XL_Z^T=I_{n_{\mathrm L}}.$

Consider a CSS code
$\mathcal{Q}_{\mathrm L}
=[[n_{\mathrm L},k_{\mathrm L},d_{\mathrm L}]]$,
referred to as the
\emph{Logical Protection Code (LPC)},
with parity-check matrices
$G_X\in\mathbb{F}_2^{s_X\times n_{\mathrm L}},
\ 
G_Z\in\mathbb{F}_2^{s_Z\times n_{\mathrm L}},$
satisfying
$G_XG_Z^T=0.$

The \emph{logically augmented CSS code}
$\mathcal{Q}_{\mathrm A}$ is defined by
the parity-check matrices
\[
H_X^{\mathrm A}=
\begin{bmatrix}
H_X\\
G_XL_X
\end{bmatrix},
\qquad
H_Z^{\mathrm A}=
\begin{bmatrix}
H_Z\\
G_ZL_Z
\end{bmatrix}.
\]

The resulting CSS code has parameters
$\mathcal{Q}_{\mathrm A}
=[[n,k-n_{\mathrm L}+k_{\mathrm L},d_{\mathrm A}]],$
where $d_{\mathrm A}$ denotes its minimum distance.
\end{definition}


\begin{example}[Repetition-Code LPC]
Consider a primary CSS code
$\mathcal{Q}_{\mathrm P}=[[n,k,d]]$.
We choose the LPC to be an $r$-qubit
repetition code with $X$-distance $d_X=r$
and $Z$-distance $d_Z=1$, where
$2\leq r\leq k$.
The LPC is defined by
\[
G_X=0,\qquad
G_Z=
\begin{bmatrix}
1&1&0&\cdots&0\\
0&1&1&\cdots&0\\
\vdots&&\ddots&\ddots&\vdots\\
0&\cdots&0&1&1
\end{bmatrix}.
\]
The repetition code protects against
$X$-type errors with distance $r$. The corresponding augmented CSS code is
defined by
\[
H_X^{\mathrm A}=H_X,\qquad
H_Z^{\mathrm A}=
\begin{bmatrix}
H_Z\\
G_ZL_Z
\end{bmatrix},
\]
where $L_Z\in\mathbb F_2^{r\times n}$
contains the selected logical $Z$ operators
of the primary code.

The resulting code has parameters
$\mathcal{Q}_{\mathrm A}
=[[n,k-r+1,d_{\mathrm A}]].$
Thus, the repetition-code LPC introduces
$r-1$ independent logical-level stabilizer
constraints without increasing the
physical block length.
\end{example}

\begin{remark}[Weight of Additional Stabilizers]
In the repetition-code example, each row of
$G_Z$ has Hamming weight two. Consequently,
each additional $Z$-type stabilizer is formed
from the product of two selected logical $Z$
operators. Its weight is at most the sum of
their weights, giving
$\mathrm{wt}\big((G_ZL_Z)_i\big)\leq 2w_Z$,
where
$w_Z=\max_j\mathrm{wt}\big((L_Z)_j\big)$.
Hence, the maximum row weight of the augmented
$Z$-type parity-check matrix satisfies
\[
w(H_Z^{\mathrm A})
\leq \max\{w(H_Z),2w_Z\}.
\]
An analogous bound holds for the $X$-type
stabilizers when the LPC imposes $X$-type
parity checks of weight two. This makes repetition codes attractive as LPCs as
each added stabilizer combines only two logical
operators, irrespective of repetition length.
General LPCs may combine more logical
operators per check and thus produce heavier
physical stabilizers, even when the parity-check
matrices of both component codes are sparse.
Therefore, the weights of the selected logical
operators and the LPC checks are important
considerations in the construction.
\end{remark}

Note that although the LA construction applies to general CSS codes, the remainder of this paper focuses on CSS-type QLDPC primary codes and the hierarchical decoding of the resulting LA codes.

\section{Hierarchical Decoding of LA QLDPC Codes}
\label{sec:decoding}

\subsection{Hierarchical Logical List Decoding (HLLD)}
\label{subsec:hlld}

We propose Hierarchical Logical List Decoding (HLLD), a BP-based framework for LA QLDPC codes that exploits the hierarchical structure of these QLDPC codes. As we will see in Section~\ref{sec:numerical}, HLLD improves decoding performance compared with applying BP directly to the full augmented parity-check matrix of the same LA QLDPC code, referred to as \textit{direct BP}. HLLD first applies BP to the primary code and invokes logical-level decoding only if this attempt fails to converge. Instead of incorporating the LPC constraints directly into BP, it uses the LPC syndrome to generate a list of admissible logical error hypotheses. Each hypothesis supplies additional constraints for a separate BP run, using the added logical redundancy to introduce structured decoding diversity. 

We next describe the decoding procedure and logical-list generation strategies. We focus on Pauli-$X$ errors as decoding of
Pauli-$Z$ errors follows analogously.
Let $e\in\mathbb F_2^n$ denote the physical
error vector. The primary and LPC syndromes
are given by
$s=H_Ze^T,\  t=G_ZL_Ze^T,$
respectively. Define the logical error vector
$u \triangleq L_Ze^T\in\mathbb F_2^{n_{\mathrm L}}.$
Although $u$ is not directly measured, the
LPC syndrome satisfies $G_Zu=t$ and thus
restricts the possible logical error
candidates to
\begin{equation}
\mathcal U(t)=
\left\{
u\in\mathbb F_2^{n_{\mathrm L}}:
G_Zu=t
\right\}.
\label{eq:admissible_logical_set}
\end{equation}

The logical-list generator selects a subset
of these candidates,
\begin{equation}
\mathcal U_L(t)=
\left\{
u^{(1)},u^{(2)},\ldots,u^{(L)}
\right\}
\subseteq\mathcal U(t),
\label{eq:logical_list}
\end{equation}
where $L$ denotes the list size. Each
hypothesis $u^{(\ell)}$ specifies additional
constraints on the physical error vector.
Accordingly, we define the conditioned parity-check matrix
and its corresponding syndrome as
\begin{equation}
H_{\mathrm C}=
\begin{bmatrix}
H_Z\\
L_Z
\end{bmatrix},
\qquad
s_{\mathrm C}^{(\ell)}=
\begin{bmatrix}
s\\
u^{(\ell)}
\end{bmatrix}.
\label{eq:conditioned_matrix}
\end{equation}
For each logical hypothesis, BP is performed
independently to obtain
\begin{equation}
\hat e^{(\ell)}
=
\operatorname{BP}
\left(
H_{\mathrm C},
s_{\mathrm C}^{(\ell)},p
\right),
\qquad \ell=1,\ldots,L.
\label{eq:conditioned_bp}
\end{equation}
The conditioned matrix $H_{\mathrm C}$ is
common to all branches, while the syndrome
$s_{\mathrm C}^{(\ell)}$ depends on the
selected logical hypothesis. Thus, the
LPC provides additional information that
allows BP to explore different logical
error configurations and generate multiple
decoding candidates.

The hierarchical procedure begins by applying
BP to the primary parity-check matrix $H_Z$
using the measured syndrome $s$. If a
syndrome-consistent estimate is obtained,
the decoder returns the corresponding
correction. Otherwise, the LPC syndrome
$t$ is measured and used to generate
$\mathcal U_L(t)$. Conditioned BP decoding
is then performed for each hypothesis in
the list, and only estimates satisfying the
conditioned syndrome constraints are retained:
\begin{equation}
\mathcal L_{\mathrm E}
=
\left\{
\hat e^{(\ell)}:
H_{\mathrm C}(\hat e^{(\ell)})^T
=s_{\mathrm C}^{(\ell)}
\right\}.
\label{eq:physical_candidate_list}
\end{equation}
The final correction is selected according
to the minimum-Hamming-weight criterion,
$\hat e=\arg\min_{\tilde e\in\mathcal L_{\mathrm E}}
\operatorname{wt}(\tilde e)$.

The conditioned BP branches can be executed
in parallel, allowing the list size to
control the tradeoff between decoding
diversity and computational cost.
The complete procedure is summarized in
Algorithm \ref{alg:hlld}. The choice of
logical-list generation strategy is
discussed in the following subsection.

\begin{algorithm}[t]
\caption{Hierarchical Logical List Decoding (HLLD)}
\label{alg:hlld}
\begin{algorithmic}[1]
\Require $H_Z$, $L_Z$, $G_Z$, primary syndrome
$s$, channel parameter $p$, list-generation
mode $\mathsf{mode}$
\Ensure Error estimate $\hat e$ or
\textsc{Failure}

\State $\hat e_{\mathrm P}
\gets\operatorname{BP}(H_Z,s,p)$

\If{$H_Z\hat e_{\mathrm P}^T=s$}
    \State \Return $\hat e_{\mathrm P}$
\EndIf

\State Measure the LPC syndrome $t$

\State $\mathcal U_L(t)\gets
\operatorname{ListGen}(G_Z, L_Z, t, \mathsf{mode})$

\State $H_{\mathrm C}\gets
\begin{bmatrix}H_Z\\L_Z\end{bmatrix}$

\State $\mathcal L_{\mathrm E}\gets\emptyset$

\For{$u^{(\ell)}\in\mathcal U_L(t)$}
    \State $s_{\mathrm C}^{(\ell)}
    \gets
    \begin{bmatrix}s\\u^{(\ell)}\end{bmatrix}$

    \State $\hat e^{(\ell)}
    \gets\operatorname{BP}
    (H_{\mathrm C},s_{\mathrm C}^{(\ell)},p)$

    \If{$H_{\mathrm C}
    (\hat e^{(\ell)})^T=s_{\mathrm C}^{(\ell)}$}
        \State $\mathcal L_{\mathrm E}
        \gets\mathcal L_{\mathrm E}
        \cup\{\hat e^{(\ell)}\}$
    \EndIf
\EndFor

\If{$\mathcal L_{\mathrm E}=\emptyset$}
    \State \Return \textsc{Failure}
\EndIf

\State \Return
$\displaystyle
\arg\min_{\tilde e\in\mathcal L_{\mathrm E}}
\operatorname{wt}(\tilde e)$

\end{algorithmic}
\end{algorithm}
\vspace{-2mm}

\subsection{Logical-List Generation Strategies}
\label{subsec:list_strategies}

The function
$\operatorname{ListGen}(G_Z,L_Z,t,\mathsf{mode})$ in Algorithm \ref{alg:hlld} determines the logical error hypotheses used
by HLLD. We consider two strategies:
maximum a posteriori (MAP)-based logical conditioning and exhaustive
logical-list decoding.

In the MAP-based strategy, only the most
probable logical error configuration
consistent with the LPC syndrome is selected.
Let $P_U(u)$ denote the probability of the
logical error vector $u=L_Ze^T$ under the
physical noise model. The MAP hypothesis is
given by
$\hat u_{\mathrm{MAP}}
=
\underset{u\in\mathcal U(t)}
{\arg\max}\;P_U(u)$
and the resulting list is
$\mathcal U_L(t)=\{\hat u_{\mathrm{MAP}}\}$,
with $L=1$. Thus, this strategy requires
only one conditioned BP execution when
the primary decoder fails to converge.

In the exhaustive strategy, all logical
error configurations consistent with the
LPC syndrome are considered:
\begin{equation}
\mathcal U_L(t)=\mathcal U(t),
\qquad
L=2^{n_{\mathrm L}-\operatorname{rank}(G_Z)}.
\label{eq:exhaustive_list}
\end{equation}
Each hypothesis defines an independent
conditioned BP decoding branch. This
strategy provides greater decoding diversity
by exploring all admissible logical
configurations.

\begin{example}[Repetition-Code LPC]
Consider the repetition-code LPC of length
$r$ introduced in the previous section.
Since $\operatorname{rank}(G_Z)=r-1$,
each LPC syndrome $t$ admits exactly two
logical error configurations:
\begin{equation}
\mathcal U(t)=
\left\{
u_0(t),\,
u_0(t)+\mathbf 1_r
\right\},
\label{eq:rep_candidates}
\end{equation}
where $u_0(t)$ is obtained by setting its
first component to zero and recursively
computing $(u_0)_{i+1}=(u_0)_i+t_i$ for $i=1,\ldots,r-1.$

The MAP-based strategy selects the more
probable of these two configurations,
resulting in a list of size one. In
contrast, exhaustive list decoding
retains both configurations, requiring
two conditioned BP branches regardless
of the repetition length.
\end{example}

\subsection{Complexity Analysis}
\label{subsec:dec_complexity}
To assess the impact of the LPC on decoding
complexity, we analyze the three main components
of HLLD: primary BP decoding, logical-list
generation, and conditioned BP decoding. Let $L$ denote the list size and $I_{\max}$ the maximum number of
BP iterations. The worst-case complexity is
$O\left(
I_{\max}|E_{\mathrm P}|
+T_{\mathrm{ListGen}}
+LI_{\max}|E_{\mathrm C}|
\right)$
where $E_{\mathrm P}$ and $E_{\mathrm C}$ denote
the edge sets of the primary and conditioned
Tanner graphs, respectively, and
$T_{\mathrm{ListGen}}$ is the cost of generating
the logical list. The latter two terms are
incurred only when primary BP fails to converge.

The complexity of list generation depends on
the LPC and the selected strategy. For MAP-based
conditioning, the logical error distribution
$P_U(u)$ can be computed using the Walsh--Hadamard
transform \cite[Ch.~5, Sec.~4]{macwilliams1977theory}
with preprocessing complexity
$O((n+n_{\mathrm L})2^{n_{\mathrm L}})$.
For larger $n_{\mathrm L}$, we instead employ
sum-product BP on a physical/logical
factor graph, defined by the constraints
$L_Ze^T=u$ and $G_Zu=t$, to approximate the
MAP logical hypothesis without computing
the full distribution.
For a repetition-code LPC of length $r$,
each syndrome admits exactly two logical
hypotheses, which can be generated in
$O(r)$ operations. Thus, exhaustive list
decoding requires at most two conditioned
BP executions, independently of $r$, while
MAP-based conditioning requires only one.
In our simulations, exact MAP probabilities
are used for $r\leq 30$, whereas BP-based
approximate MAP estimation is employed
for $r>30$.

\section{Simulation Results}
\label{sec:numerical}

In this section, we evaluate the proposed
framework of LA QLDPC code constructions and HLLD
over a binary symmetric channel with independent
Pauli-$X$ errors of probability $p_x$. Each simulation
point is obtained after at least 100 decoding
failures. All BP decoders employ sum-product
decoding with parallel (flooding) scheduling
and a maximum of $I_{\max}=50$ iterations.
For numerical stability, the variable-to-check
messages are clipped to $[-K,K]$, with $K=25$.
For BP-OSD, we use order-zero OSD (OSD-0)
implemented using the \texttt{ldpc} Python library
\cite{Roffe_LDPC_Python_tools_2022,Roffe_2020}.
To evaluate the benefits of the joint framework,
we consider two types of comparisons. First,
we compare the performance of LA codes with
that of their underlying primary codes to
assess the effect of introducing logical-level
redundancy. Second, we compare HLLD with
conventional decoding of the same LA code
to isolate the performance gains arising
from the hierarchical decoding strategy.

Throughout this section, we use
HLLD-MAP for MAP-based logical-list generation,
which considers the most probable logical
hypothesis, and HLLD-FULL for exhaustive
list generation, which considers all
syndrome-consistent logical hypotheses.
For repetition-code LPCs, HLLD-FULL requires
a list size of two, independently of the
repetition length. We compare the proposed HLLD strategies with
BP and BP-OSD decoding of the primary and LA codes. We also use direct BP and direct BP-OSD to refer to decoding using the full augmented
parity-check matrix, whereas HLLD-MAP
and HLLD-FULL exploit the hierarchical
structure of the LA code. For HLLD with BP-OSD, the hierarchical stage is triggered by primary BP nonconvergence,
without OSD post-processing. OSD is applied
only within the conditioned decoding branches.

We first demonstrate a trade-off between
encoding rate and logical error-rate performance
by varying the repetition LPC length in a certain example.
Table \ref{tab:repetition} presents the results
for the $[[2610,744]]$ lifted-product (LP) code \cite{cain2026shor} at $p_x=0.02$
using HLLD-FULL. As the repetition length
increases, the additional logical-level
redundancy leads to a robust improvement in error-correction
performance at the cost of a moderate rate
reduction. In particular, increasing the LPC
length to $r=40$ reduces the logical error rate
by a factor of $2.5$, while decreasing the
encoding rate by only $5.24\%$. These results
demonstrate how the proposed LA construction
provides a flexible means of exploiting
logical-level redundancy to improve error
protection without increasing the physical
block length.


Moreover, across various QLDPC code families, the proposed
LA-HLLD framework achieves substantial improvements
in logical error rate compared with conventional
decoding of the primary and LA codes.
Fig. \ref{fig:fer_loglogB1} shows the results
for the $[[2610,744]]$ LP code \cite{cain2026shor}, where a repetition
LPC of length $r=50$ yields the LA-LP code
$[[2610,695]]$, with $6.59\%$ rate reduction.
At $p_x=0.015$, HLLD-MAP and HLLD-FULL achieve
approximately $62\%$ and $75\%$ reductions
in logical error rate, respectively, relative
to direct BP.
Fig. \ref{fig:fer_loglog144_clipped2}
presents the results for the $[[144,12,12]]$
Bivariate Bicycle (BB) code \cite{Bravyi_2024} with a repetition LPC of length $r=3$,
corresponding to $16.67\%$ rate reduction.
At $p_x=0.015$, HLLD-MAP and HLLD-FULL reduce
the logical error rate by approximately $37\%$
and $69\%$, respectively, compared with direct
BP. The improvements also extend to BP-OSD-based
decoding in the low-error regime, demonstrating
the benefits of exploiting logical-level
constraints through hierarchical list decoding.
Finally, Fig. \ref{fig:fer_loglogMitten} presents the results
for the $[[780,156]]$ mitten code \cite{bhardwaj2026high}, where a repetition
LPC of length $r=10$ yields the LA code $[[780,147]]$,
corresponding to only $5.77\%$ rate reduction.
At $p_x=0.02$, HLLD-MAP and HLLD-FULL reduce
the logical error rate by approximately $28\%$
and $81\%$, respectively, compared with direct
BP decoding of the same LA code.

\begin{table}[t!]
\centering
\caption{Rate-performance trade-off for LA codes
constructed from the $[[2610,744]]$ lifted-product (LP) code \cite{cain2026shor}
with varying repetition length $r$,
using HLLD-FULL at $p_x=0.02$.
Here, $P_L^{(r)}$ denotes the logical error rate
for repetition length $r$.}

\label{tab:repetition}
\begin{tabular}{lccccc}
\toprule
Repetition length $r$ & 0 (Primary) & 10 & 20 & 30 & 40 \\
\midrule
Rate reduction (\%) & 0 & 1.21 & 2.55 & 3.90 & 5.24 \\
$P_L^{(0)}/P_L^{(r)}$ & 1.00 & 1.53 & 1.74 & 2.20 & 2.50 \\
\bottomrule
\end{tabular}
\vspace{-1mm}

\end{table}


\definecolor{colorDirectBP}{HTML}{8C8C8C}
\definecolor{colorLocal}{HTML}{FF7F0E}
\definecolor{colorTriggered}{HTML}{1F77B4}
\definecolor{colorTriggeredAll}{HTML}{6A0DAD}

\begin{figure}[t]
    \centering
  \resizebox{\columnwidth}{!}{%
\begin{tikzpicture}
\begin{loglogaxis}[
    scale only axis,
    width  = 8.4cm,
    height = 5.6cm,
        xlabel = {Physical Error Rate $p_x$},
        ylabel = {Logical Error Rate},
        xlabel style     = {font=\small},
        ylabel style     = {font=\small},
        tick label style = {font=\small},
        xmin = 0.014, xmax = 0.052,
        ymin = 1e-4,  ymax = 1.2,
        xtick       = {0.015, 0.02, 0.025, 0.03, 0.04, 0.05},
        xticklabels = {$0.015$, $0.02$, $0.025$, $0.03$, $0.04$, $0.05$},
        xticklabel style = {rotate=25, anchor=north east, font=\small},
        xtick align      = inside,
        ytick align      = inside,
        enlarge x limits = 0.02,
        enlarge y limits = 0.02,
        clip             = true,
        grid             = both,
        grid style       = {line width=0.4pt, black!25},
        minor grid style = {line width=0.3pt, black!15},
        minor tick num   = 1,
        axis line style  = {line width=0.8pt},
        legend style = {
            at           = {(0.99, 0.01)},
            anchor       = south east,
            font         = \footnotesize,
            fill         = white,
            fill opacity = 0.85,
            text opacity = 1,
            draw         = black!30,
            rounded corners = 2pt,
            cells        = {anchor=west},
            column sep   = 2pt,
            row sep      = 1pt,
            inner sep    = 4pt,
        },
        legend columns    = 1,
        legend cell align = left,
    ]


    \addplot[forget plot, color=colorLocal, mark=triangle*, mark size=3.5pt,
             line width=1.4pt, solid,
             mark options={solid}] coordinates {
        (0.015,3.961e-04)(0.02,2.531e-03)(0.025,1.065e-02)(0.03,4.032e-02)
        (0.04,2.632e-01)(0.05,7.972e-01)
    };
    \addlegendimage{color=colorLocal, mark=triangle*, mark size=3.5pt,
             line width=1.4pt, solid, mark options={solid}}
    \addlegendentry{BP, LP\,[[2610, 744]]}
    
    \addplot[color=colorDirectBP, mark=square*, mark size=3pt,
             line width=1.4pt, solid] coordinates {
        (0.015,6.301e-04)(0.02,2.121e-03)(0.025,7.291e-03)(0.03,2.845e-02)
        (0.04,1.807e-01)(0.05,6.993e-01)
    };
     \addlegendentry{BP, LA-LP\,[[2610, 695]]}

    \addplot[color=colorTriggered, mark=*, mark size=3pt,
             line width=1.4pt, solid] coordinates {
        (0.015,2.402e-04)(0.02,1.681e-03)(0.025,8.031e-03)(0.03,3.136e-02)
        (0.04,2.087e-01)(0.05,7.483e-01)
    };
    \addlegendentry{HLLD-MAP, LA-LP\,[[2610, 695]]}

    \addplot[color=colorTriggeredAll, mark=star, mark size=4.5pt,
             mark options={fill=colorTriggeredAll, line width=0.8pt},
             line width=1.4pt, solid] coordinates {
        (0.015,1.560e-04)(0.02,1.025e-03)(0.025,5.696e-03)(0.03,2.240e-02)
        (0.04,1.558e-01)(0.05,7.133e-01)
    };
    \addlegendentry{HLLD-FULL, LA-LP\,[[2610, 695]]}

    \end{loglogaxis}
    \end{tikzpicture}}%
    
\caption{Decoding performance of the logically
augmented $[[2610,695]]$ code constructed from
the $[[2610,744]]$ LP code \cite{cain2026shor}
using a repetition LPC of length $r=50$, under direct BP and the proposed HLLD-MAP and HLLD-FULL.}

    \label{fig:fer_loglogB1}
    \vspace{-2mm}

\end{figure}
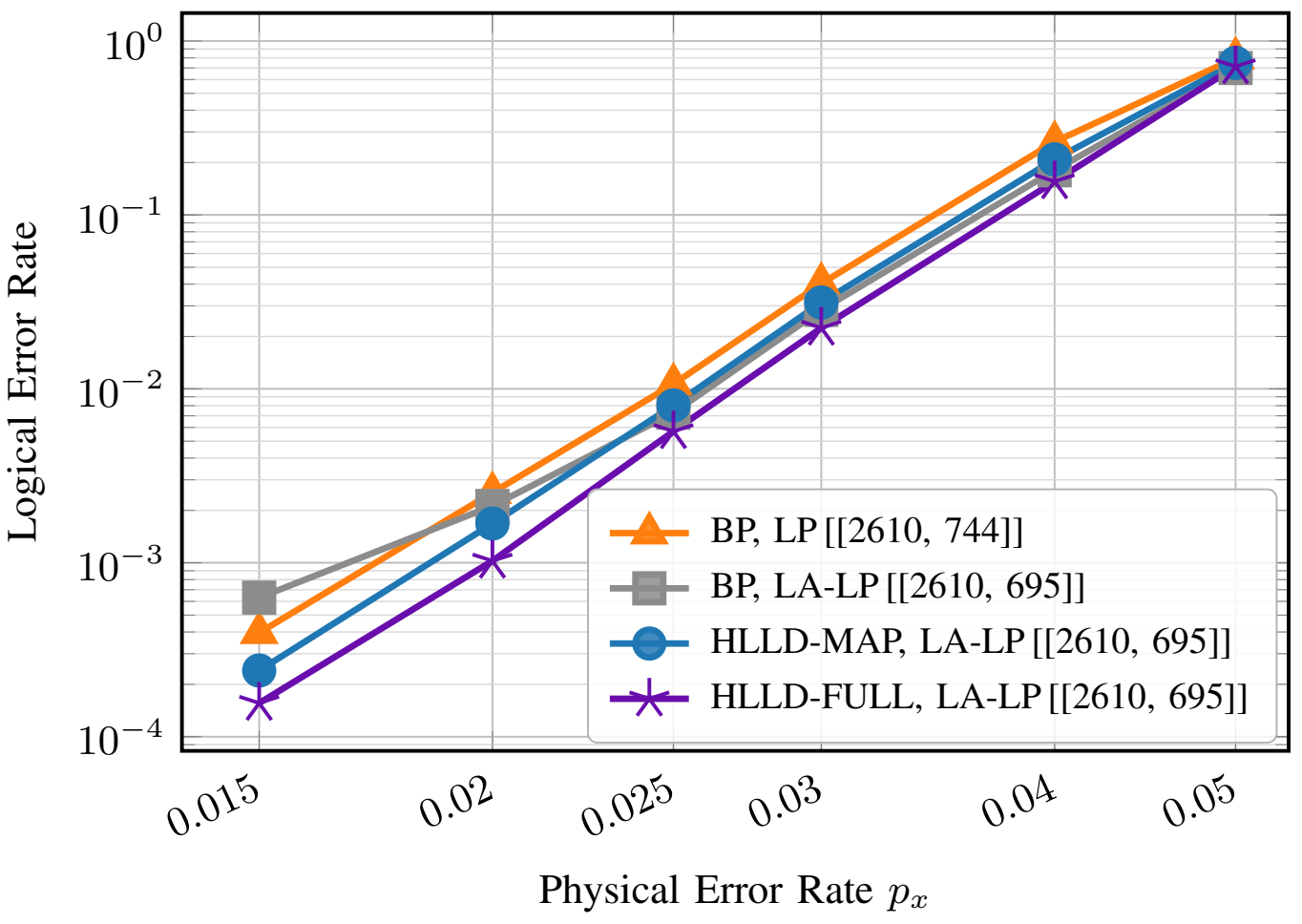

\definecolor{colorDirect}{HTML}{8C8C8C}
\definecolor{colorLocal}{HTML}{FF7F0E}
\definecolor{colorTriggered}{HTML}{1F77B4}
\definecolor{colorTriggeredAll}{HTML}{6A0DAD}

\begin{figure*}[t]
    \centering
    \resizebox{0.48\textwidth}{!}{%
\begin{tikzpicture}
\begin{loglogaxis}[
    scale only axis,
    width  = 8.4cm,
    height = 5.6cm,
        xlabel = {Physical Error Rate $p_x$},
        ylabel = {Logical Error Rate},
        xlabel style     = {font=\small},
        ylabel style     = {font=\small},
        tick label style = {font=\small},
        xmin = 0.0095, xmax = 0.105,
        ymin = 2e-5,  ymax = 1.2,
        xtick       = {0.01, 0.015, 0.02, 0.03, 0.05, 0.07, 0.1},
        xticklabels = {$0.01$, $0.015$, $0.02$, $0.03$, $0.05$, $0.07$, $0.10$},
        xticklabel style = {rotate=25, anchor=north east, font=\small},
        xtick align      = inside,
        ytick align      = inside,
        enlarge x limits = 0.02,
        enlarge y limits = 0.02,
        clip             = true,
        grid             = both,
        grid style       = {line width=0.4pt, black!25},
        minor grid style = {line width=0.3pt, black!15},
        minor tick num   = 1,
        axis line style  = {line width=0.8pt},
        legend style = {
            at           = {(0.99, 0.01)},
            anchor       = south east,
            font         = \footnotesize,
            fill         = white,
            fill opacity = 0.85,
            text opacity = 1,
            draw         = black!30,
            rounded corners = 2pt,
            cells        = {anchor=west},
            column sep   = 2pt,
            row sep      = 1pt,
            inner sep    = 4pt,
        },
        legend columns    = 1,
        legend cell align = left,
        title = {(a) BP},
        title style = {font=\small},
    ]


    \addlegendimage{color=colorLocal, mark=triangle*, mark size=3.5pt,
             line width=1.4pt, solid, mark options={solid}}
    \addlegendentry{BP, BB\,[[144,\,12]]}
    \addplot[forget plot, color=colorLocal, mark=triangle*, mark size=3.5pt,
             line width=1.4pt, solid,
             mark options={solid}] coordinates {
        (0.01,5.617e-04)(0.015,1.239e-03)(0.02,3.121e-03)(0.025,5.570e-03)(0.03,1.042e-02)(0.04,3.128e-02)
        (0.05,8.196e-02)(0.06,1.562e-01)(0.07,2.716e-01)(0.08,4.300e-01)(0.09,5.533e-01)(0.1,7.653e-01)
    };

    \addplot[color=colorDirect, mark=square*, mark size=3pt,
             line width=1.4pt, solid] coordinates {
        (0.01,2.440e-04)(0.015,8.094e-04)(0.02,1.692e-03)(0.025,3.816e-03)(0.03,6.774e-03)(0.04,2.259e-02)
        (0.05,6.113e-02)(0.06,1.256e-01)(0.07,2.379e-01)(0.08,3.955e-01)(0.09,5.262e-01)(0.1,7.413e-01)
    };
    \addlegendentry{BP, LA-BB\,[[144,\,10]]}
    
    \addplot[color=colorTriggered, mark=*, mark size=3pt,
             line width=1.4pt, solid] coordinates {
        (0.01,2.083e-04)(0.015,5.079e-04)(0.02,1.336e-03)(0.025,2.871e-03)(0.03,6.007e-03)(0.04,2.234e-02)
        (0.05,6.416e-02)(0.06,1.305e-01)(0.07,2.336e-01)(0.08,4.009e-01)(0.09,5.374e-01)(0.1,7.333e-01)
    };
    \addlegendentry{HLLD-MAP, LA-BB\,[[144,\,10]]}

    \addplot[color=colorTriggeredAll, mark=star, mark size=4.5pt,
             line width=1.4pt, solid,
             mark options={solid, line width=0.8pt}] coordinates {
        (0.01,1.099e-04)(0.015,2.502e-04)(0.02,7.485e-04)(0.025,1.674e-03)(0.03,3.582e-03)(0.04,1.610e-02)
        (0.05,4.995e-02)(0.06,1.109e-01)(0.07,2.177e-01)(0.08,3.685e-01)(0.09,5.072e-01)(0.1,7.173e-01)
    };
    \addlegendentry{HLLD-FULL, LA-BB\,[[144,\,10]]}

    \end{loglogaxis}
    \end{tikzpicture}}%
    \hfill
    \resizebox{0.48\textwidth}{!}{%
\begin{tikzpicture}
\begin{loglogaxis}[
    scale only axis,
    width  = 8.4cm,
    height = 5.6cm,
        xlabel = {Physical Error Rate $p_x$},
        ylabel = {Logical Error Rate},
        xlabel style     = {font=\small},
        ylabel style     = {font=\small},
        tick label style = {font=\small},
        xmin = 0.0095, xmax = 0.105,
        ymin = 2e-5,  ymax = 1.2,
        xtick       = {0.01, 0.015, 0.02, 0.03, 0.05, 0.07, 0.1},
        xticklabels = {$0.01$, $0.015$, $0.02$, $0.03$, $0.05$, $0.07$, $0.10$},
        xticklabel style = {rotate=25, anchor=north east, font=\small},
        xtick align      = inside,
        ytick align      = inside,
        enlarge x limits = 0.02,
        enlarge y limits = 0.02,
        clip             = true,
        grid             = both,
        grid style       = {line width=0.4pt, black!25},
        minor grid style = {line width=0.3pt, black!15},
        minor tick num   = 1,
        axis line style  = {line width=0.8pt},
        legend style = {
            at           = {(0.99, 0.01)},
            anchor       = south east,
            font         = \footnotesize,
            fill         = white,
            fill opacity = 0.85,
            text opacity = 1,
            draw         = black!30,
            rounded corners = 2pt,
            cells        = {anchor=west},
            column sep   = 2pt,
            row sep      = 1pt,
            inner sep    = 4pt,
        },
        legend columns    = 1,
        legend cell align = left,
        title = {(b) BP-OSD},
        title style = {font=\small},
    ]


    \addlegendimage{color=colorLocal, mark=triangle, mark size=3.5pt,
             line width=1.4pt, solid, mark options={solid}}
    \addlegendentry{BP-OSD, BB\,[[144,\,12]]}
    \addplot[forget plot, color=colorLocal, mark=triangle, mark size=3.5pt,
             line width=1.4pt, solid,
             mark options={solid}] coordinates {
        (0.01,2.666e-04)(0.015,8.082e-04)(0.02,1.557e-03)(0.025,2.857e-03)(0.03,4.771e-03)(0.04,1.580e-02)
        (0.05,3.629e-02)(0.06,8.790e-02)(0.07,1.637e-01)(0.08,2.780e-01)(0.09,4.022e-01)(0.1,5.973e-01)
    };

    \addplot[color=colorDirect, mark=square, mark size=3pt,
             line width=1.4pt, solid,
             mark options={solid}] coordinates {
        (0.01,3.435e-04)(0.015,1.086e-03)(0.02,2.126e-03)(0.025,3.968e-03)(0.03,6.523e-03)(0.04,1.452e-02)
        (0.05,2.856e-02)(0.06,6.248e-02)(0.07,1.226e-01)(0.08,2.166e-01)(0.09,3.243e-01)(0.1,5.333e-01)
    };
    \addlegendentry{BP-OSD, LA-BB\,[[144,\,10]]}

    \addplot[color=colorTriggered, mark=o, mark size=3pt,
             line width=1.4pt, solid,
             mark options={solid}] coordinates {
        (0.01,2.061e-04)(0.015,5.542e-04)(0.02,1.370e-03)(0.025,2.626e-03)(0.03,5.804e-03)(0.04,1.840e-02)
        (0.05,5.092e-02)(0.06,1.017e-01)(0.07,1.833e-01)(0.08,3.028e-01)(0.09,4.165e-01)(0.1,6.267e-01)
    };
    \addlegendentry{HLLD-MAP-OSD, LA-BB\,[[144,\,10]]}

    \addplot[color=colorTriggeredAll, mark=pentagon, mark size=3.5pt,
             line width=1.4pt, solid,
             mark options={solid}] coordinates {
        (0.01,1.270e-04)(0.015,3.603e-04)(0.02,7.785e-04)(0.025,1.443e-03)(0.03,3.129e-03)(0.04,1.022e-02)
        (0.05,2.760e-02)(0.06,6.126e-02)(0.07,1.245e-01)(0.08,2.155e-01)(0.09,3.180e-01)(0.1,5.360e-01)
    };
    \addlegendentry{HLLD-FULL-OSD, LA-BB\,[[144,\,10]]}

    \end{loglogaxis}
    \end{tikzpicture}}%
    
\caption{Decoding performance of the logically
augmented code constructed from the
$[[144,12,12]]$ bivariate bicycle (BB) code
\cite{Bravyi_2024} using a repetition LPC
of length $r=3$: (a) BP and (b) BP-OSD,
comparing direct decoding with the proposed
HLLD-MAP and HLLD-FULL strategies.}

    \label{fig:fer_loglog144_clipped2}
    \vspace{-2mm}
\end{figure*}
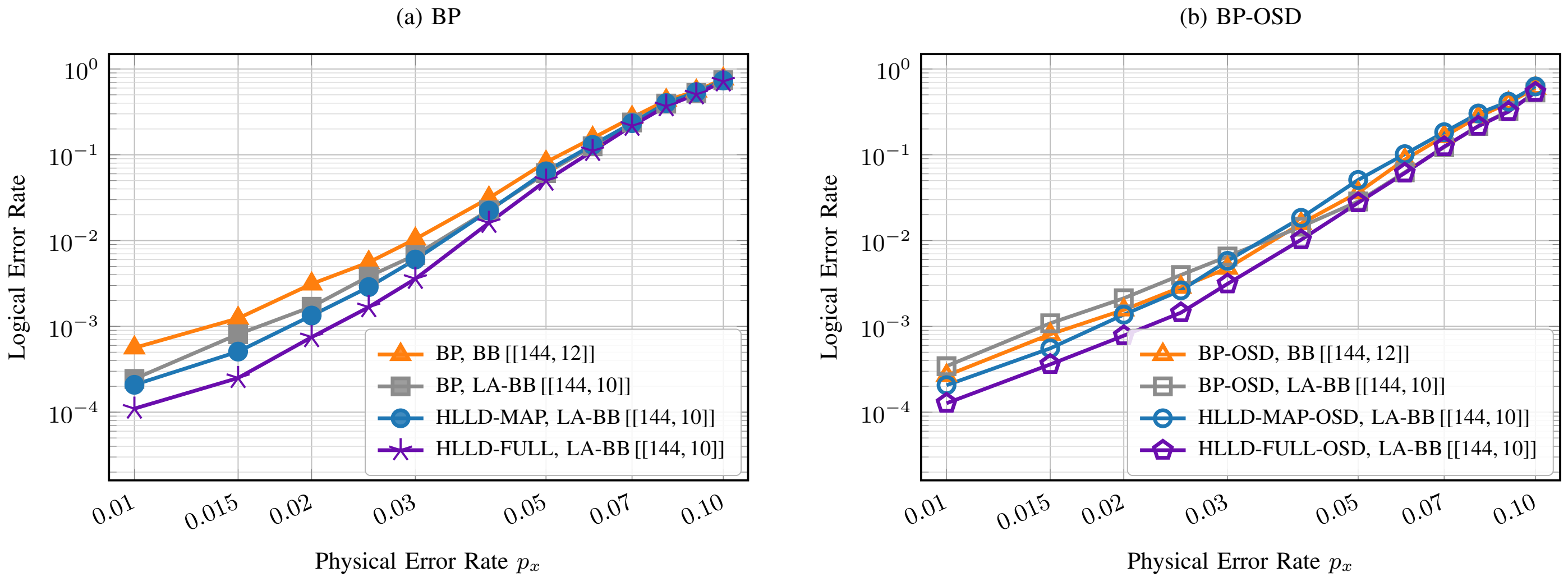

\definecolor{colorDirectBP}{HTML}{8C8C8C}
\definecolor{colorLocal}{HTML}{FF7F0E}
\definecolor{colorTriggered}{HTML}{1F77B4}
\definecolor{colorTriggeredAll}{HTML}{6A0DAD}

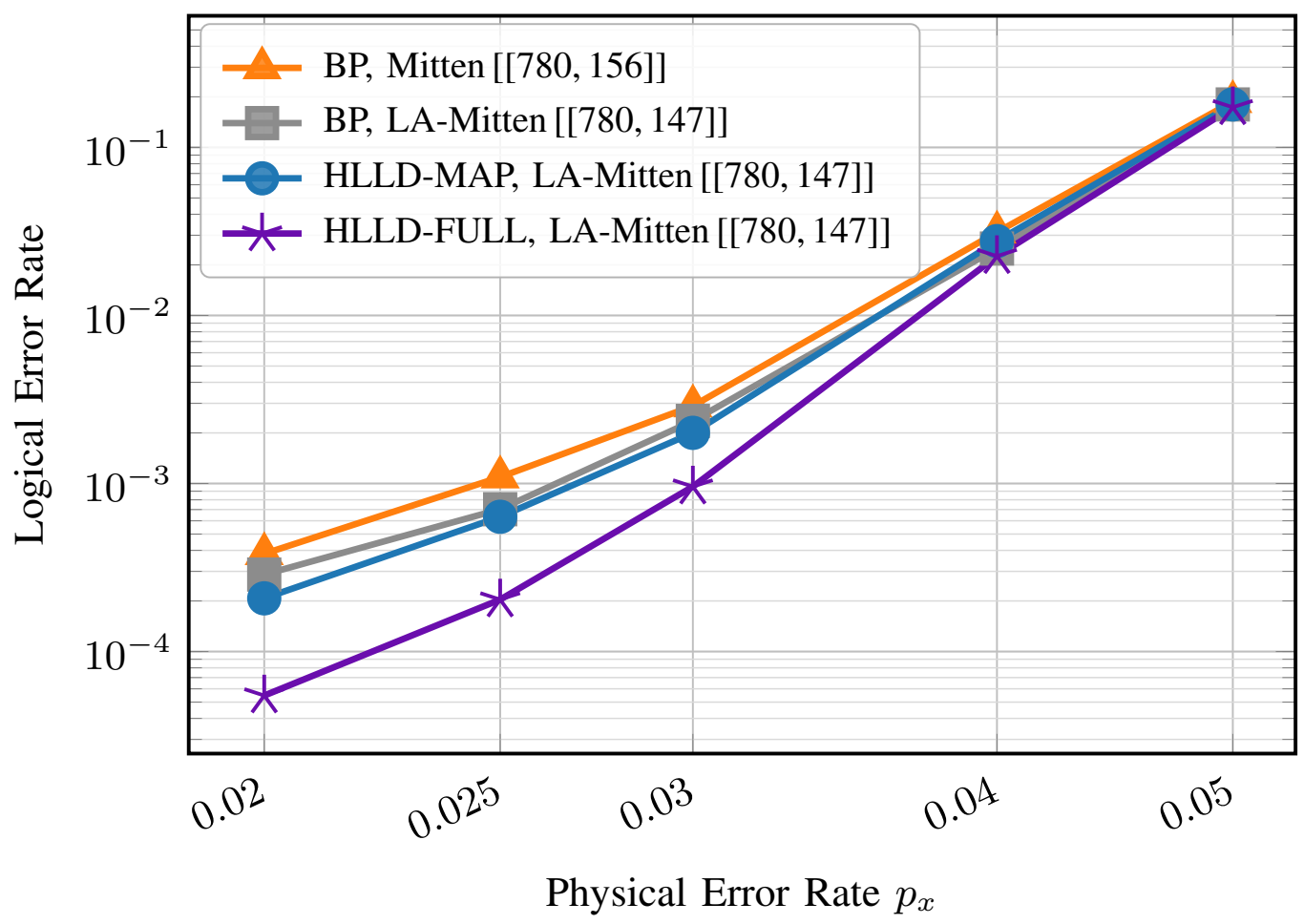
\begin{figure}[t]
    \centering
    \resizebox{\columnwidth}{!}{%
\begin{tikzpicture}
\begin{loglogaxis}[
    scale only axis,
    width  = 8.4cm,
    height = 5.6cm,
        xlabel = {Physical Error Rate $p_x$},
        ylabel = {Logical Error Rate},
        xlabel style     = {font=\small},
        ylabel style     = {font=\small},
        tick label style = {font=\small},
        xmin = 0.019, xmax = 0.052,
        ymin = 3e-5,  ymax = 0.5,
        xtick       = {0.02, 0.025, 0.03, 0.04, 0.05},
        xticklabels = {$0.02$, $0.025$, $0.03$, $0.04$, $0.05$},
        xticklabel style = {rotate=25, anchor=north east, font=\small},
        xtick align      = inside,
        ytick align      = inside,
        enlarge x limits = 0.02,
        enlarge y limits = 0.02,
        clip             = true,
        grid             = both,
        grid style       = {line width=0.4pt, black!25},
        minor grid style = {line width=0.3pt, black!15},
        minor tick num   = 1,
        axis line style  = {line width=0.8pt},
        legend style = {
            at           = {(0.01, 0.99)},
            anchor       = north west,
            font         = \footnotesize,
            fill         = white,
            fill opacity = 0.85,
            text opacity = 1,
            draw         = black!30,
            rounded corners = 2pt,
            cells        = {anchor=west},
            column sep   = 2pt,
            row sep      = 1pt,
            inner sep    = 4pt,
        },
        legend columns    = 1,
        legend cell align = left,
    ]


    \addlegendimage{color=colorLocal, mark=triangle*, mark size=3.5pt,
             line width=1.4pt, solid, mark options={solid}}
    \addlegendentry{BP, Mitten\,[[780,\,156]]}
    \addplot[forget plot, color=colorLocal, mark=triangle*, mark size=3.5pt,
             line width=1.4pt, solid,
             mark options={solid}] coordinates {
        (0.02,3.819e-04)(0.025,1.092e-03)(0.03,2.893e-03)(0.04,3.131e-02)(0.05,1.872e-01)
    };

    \addplot[color=colorDirectBP, mark=square*, mark size=3pt,
             line width=1.4pt, solid] coordinates {
        (0.02,2.872e-04)(0.025,7.030e-04)(0.03,2.357e-03)(0.04,2.501e-02)(0.05,1.802e-01)
    };
    \addlegendentry{BP, LA-Mitten\,[[780,\,147]]}

    \addplot[color=colorTriggered, mark=*, mark size=3pt,
             line width=1.4pt, solid] coordinates {
        (0.02,2.073e-04)(0.025,6.337e-04)(0.03,2.002e-03)(0.04,2.771e-02)(0.05,1.802e-01)
    };
    \addlegendentry{HLLD-MAP, LA-Mitten\,[[780,\,147]]}

    \addplot[color=colorTriggeredAll, mark=star, mark size=4.5pt,
             line width=1.4pt, solid,
             mark options={fill=colorTriggeredAll, line width=0.8pt}] coordinates {
        (0.02,5.471e-05)(0.025,2.038e-04)(0.03,9.580e-04)(0.04,2.253e-02)(0.05,1.733e-01)
    };
    \addlegendentry{HLLD-FULL, LA-Mitten\,[[780,\,147]]}

    \end{loglogaxis}
    \end{tikzpicture}}%
    
\caption{Decoding performance of the logically
augmented $[[780,147]]$ code constructed from
the $[[780,156]]$ mitten code \cite{bhardwaj2026high}
using a repetition LPC of length $r=10$, under direct BP and the proposed HLLD-MAP and HLLD-FULL.}

    \label{fig:fer_loglogMitten}
    \vspace{-2mm}

\end{figure}

\section{Conclusion}
\label{sec:conclusion}

We proposed a joint construction and decoding framework
for logically augmented QLDPC codes. The framework introduces additional
stabilizer constraints on selected logical qubits, enabling flexible rate and asymmetric error protection while preserving the physical block length and original stabilizer generators. On the decoding side, HLLD exploits the resulting syndrome information to generate logical error hypotheses
and guide conditioned BP decoding when primary BP fails to converge. Numerical results demonstrated a tunable rate--performance trade-off and lower logical error rates than direct BP decoding of the same LA codes.

Future work includes extending the study to additional
QLDPC code families
\cite{PhysRevA.88.012311,
wang2022distanceboundsgeneralizedbicycle,
rabeti2026univariate},
developing efficient MAP-based hypothesis selection
for larger codes, optimizing the LPC repetition length,
and evaluating the framework under biased noise models.

\vspace{-1.5mm}

\bibliographystyle{IEEEtran}
{\footnotesize \bibliography{reffff}}
\end{document}